# Atomic Topology and Magnetic Microstructure of Highly Mobile Type I and Supermobile Type II Twin Boundaries in 10M Ni–Mn–Ga Single Crystal

Ladislav Straka[1], Marek Vronka[1,*], Jan Maňák[1], Petr Veřtát[1], Hanuš Seiner[2], Oleg Heczko[1]

[1]*FZU – Institute of Physics of the Czech Academy of Sciences, Prague 8, 182 00, Czechia*

[2] *Institute of Thermomechanics of the Czech Academy of Sciences, Prague 8, 182 00, Czechia*

**Abstract**

The atomic topology and magnetic microstructure of individual, highly mobile Type I and Type II twin boundaries in 10M Ni–Mn–Ga martensite were investigated by transmission electron microscopy (TEM). The twin boundaries established in a bulk single crystal showed twinning stresses of ~1 MPa for Type I and ~0.1 MPa for Type II twin boundaries. TEM lamellae with a (010) cross-section, their *c*-axis (easy-magnetization direction) lying in-plane, were prepared by focused ion-beam milling, each containing a single twin boundary of specific type. High-resolution TEM confirmed an atomically sharp Type I twin boundary oriented along the rational (101) plane. The Type II boundary was also atomically sharp, apart from occasional single-atomic-plane steps. This contrasts with previous suggestions of its diffuse nature. Lorentz TEM showed 180° domain walls within martensite variants. The magnetic induction reorients sharply on both twin boundaries, forming 90°-like magnetic domain walls that follow the *c*-axis easy-magnetization direction.

*\* Corresponding author, email: vronka@fzu.cz, phone: +420266052870*

The extraordinarily high mobility of martensite twin boundaries is fundamental to magnetic shape-memory alloys and relevant to many other functional-material systems. Magnetic-field–induced strain (MFIS) arises when a field drives magnetically induced reorientation (MIR) of martensite variants, a process enabled by the motion of martensite twin boundaries (TBs)—highly mobile internal interfaces that switch one martensite variant into another. Under slow-loading (quasistatic) conditions, twin-boundary mobility is quantified by the twinning stress [1–4].

Within Ni–Mn–Ga five-layered modulated (10M) martensite, the prototype magnetic shape memory material, two principal types of twin boundary exist—Type I and Type II. Type I TBs typically exhibit twinning stresses around 1 MPa or higher, whereas Type II TBs can move under stresses approaching a mere 0.1 MPa [4–7]. This difference is accompanied by markedly distinct temperature dependencies: Type II TBs retain their high mobility even at temperatures as low as 2 K. This contrasts sharply with the temperature-dependent behavior (~0.04 MPa/K) observed for Type I TBs, which hinders the MIR at low temperatures [8–10]. The extraordinarily low stress required for motion of Type II TBs, complemented by their weak temperature dependence, has justifiably led to Type II boundaries being labeled *supermobile* [11,12].

The origin of the supermobility in Type II TBs remains a puzzle and is the subject of intense research. It has been suggested that it may be related to the extreme elastic anisotropy of Ni–Mn–Ga [12]. However, the microstructure and atomistic topology of the boundaries are also widely expected to play a significant role [13–16]. The topology is a crucial starting point for mechanistic considerations: Type I TBs, with their rational {101} plane, directly imply an atomically sharp boundary. Conversely, Type II TBs with irrational indices close to the {10 1 10} were initially anticipated to form broad boundaries [5,6]. However, direct observation of these boundaries using TEM is challenging due to their high mobility under stress and, particularly, in an external magnetic field.

Early investigations assumed that Type II TBs were broad and diffuse, ensuring minimal interaction with lattice defects that might impede their mobility [17,18]. This concept was supported by the observations of Matsuda *et al.* [19], who demonstrated a Type II TB spanning tens of atomic planes. Conversely, a more recent model proposed by Shilo *et al.* [16] considered a faceted topology of Type II TBs, i.e., composed of low-index facet planes separated by steps, a characteristic typical of interfaces formed on irrational planes. In this context, the high mobility of Type II TBs was attributed to a significantly reduced requirement for thermal nucleation of defects in a faceted—and thus already defected—interface, compared to defect-free atomically sharp Type I TB.

To directly address the long-standing debate and provide clear atomic-scale evidence, in this study, we combine transmission electron microscopy (TEM), Lorentz TEM (LTEM), and high-resolution TEM (HRTEM) methods to examine the atomic structure of Type I and Type II TBs. In contrast to previous work, where twin boundaries were studied in polycrystalline material with undefined twin mobility, our study focuses on the individual highly mobile twin boundaries prepared in a single-crystalline bulk sample. The structural examination was conducted across the (010) cut. In this orientation, the *c*-axis is in the lamella plane for both variants separated by a twin boundary, and the two variants have the same energy in a magnetic field applied out of the plane. Thus, it can be anticipated that, although a twin boundary is highly mobile, it will not move within the strong magnetic field of the TEM objective lens.

The material investigated was a bulk single crystal of $Ni_{50}Mn_{28}Ga_{22}$ obtained from Adaptamat Ltd., Finland. The crystal was cut into a parallelepiped sample with dimensions of approximately 0.9 × 2.3 × 20 $mm^3$ with its faces parallel to the {100} crystallographic planes of the parent austenite phase. At room temperature, the sample was confirmed by X-ray diffraction to exhibit a 10M martensite structure. Lattice parameters of this phase, determined from detailed two-dimensional reciprocal space maps of strong fundamental reflections and expressed in pseudo-cubic coordinates (keeping the orientation of the axes similar to the cubic austenite phase), were established as *a* = 5.965(3) Å, *b* = 5.952(3) Å, *c* = 5.585(1) Å, *γ* = 90.36(5)°. The slightly increased uncertainty in lattice parameters originates from the slightly diffuse character and overlap of the (400) and (040) reflections, possibly due to a slight incommensurability of the structural modulation [20].

Well-separated Type I and Type II twin boundaries were produced by repeated mechanical and magnetic loading along the principal sample axes. The individual boundaries were visible at the macroscopic scale. Figure 1a presents photographs of the sample containing both TB types near its centre. The Type I TB twinning plane is of the {101} orientation, whereas the Type II TB is inclined by about 4° from this plane, corresponding to a plane with Miller indices approximately {10 1 10}. Consequently, on a {100}-cut surface of the sample (oriented with [001] and [100] directions in-plane), the observed twin boundary traces showed no inclination for the Type I TB from the [100] direction, but an inclination of about 6° for the Type II TB trace.

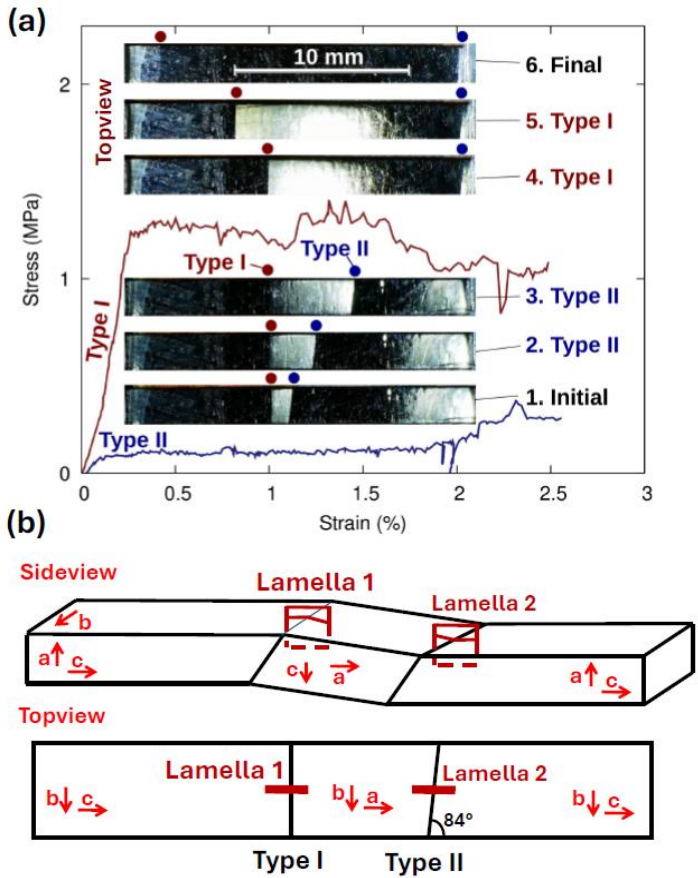

**Figure 1.** Macroscopic characterization of twin boundaries and TEM lamellae preparation. **(a)** Stress–strain curves corresponding to the motion of Type II and Type I twin boundaries and photographs of the single-crystal sample with the moving boundaries (top view): 1: initial state; 2,3: propagation of the Type II twin boundary; 4,5,6: subsequent propagation of a Type I twin boundary and the final state. The approximate positions of the Type I (red dots) and Type II (blue dots) boundaries are indicated above the micrographs. **(b)** Schematic illustration of the orientation of the TEM lamellae containing Type I and Type II twin boundaries, lifted from the sample using focused Ga-ion FIB. The lattice orientation is shown in the side and top views of the sample (disregarding the *a/b* twinning).

Both boundaries were characterized by twinning stress measurements at room temperature. The compression stressstrain curves are shown in Fig. 1(a) together with photographs documenting the propagation of boundaries. Under load, the inclined twin boundary propagated first, with a distinct stress plateau indicating a very low twinning stress of about 0.1 MPa, confirming it as Type II. After this boundary reached the end of the sample, the non-inclined boundary propagated in the opposite direction under a stress exceeding 1 MPa, confirming it as Type I [8].

Prior to this mechanical testing, site-specific Ga-ion FIB milling was used to lift out several thin lamellae (approx. 30 nm × 10 μm × 10 μm) from the sample, each containing the individual Type I or Type II boundary, as illustrated in Fig. 1b. Each lamella contained the long (*a* or *b*) and short (*c*) axes in its plane, with one of the long axes pointing out-of-plane, i.e., the c-axis was in the foil plane for both variants. Although 10M martensite typically forms *a/b* laminates, for clarity, we refer here to the in-plane long axis always as "*a*" and the out-of-plane long axis always as "*b*" (note also that *a* and *b* are nearly equal, *a/b* = 1.002).

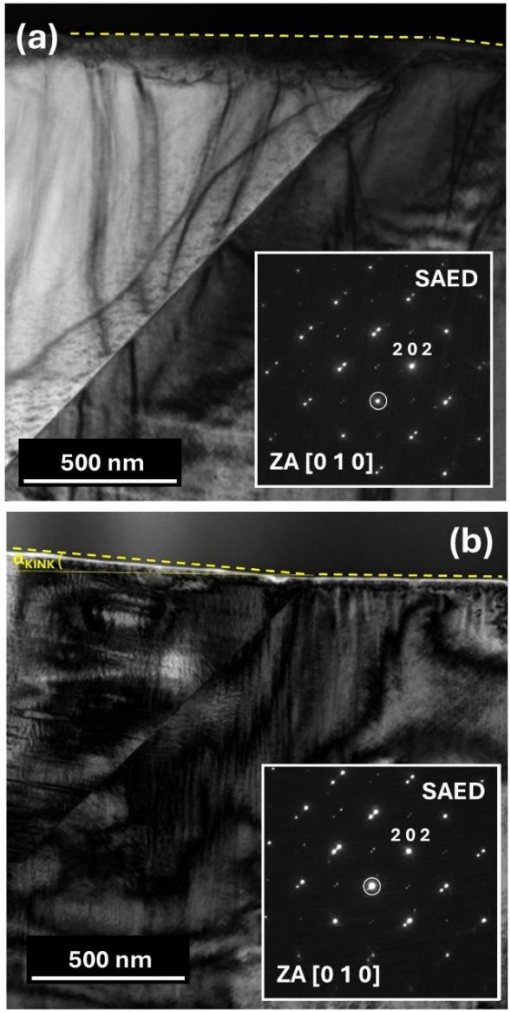

**Figure 2.** Bright field TEM micrographs with their corresponding SAED patterns (insets). The boundary is oriented at approximately 45° to the lamella edge: **(a)** Type I twin boundary, zone axis [010]; **(b)** Type II twin boundary, zone axis [010]. The yellow dashed line indicates the lamella surface, demonstrating the surface kinking due to *a/c* twinning, with $\alpha_{KINK}$ = 90 - $2arctan(c/a)$ = 3.77°.

The two types of boundaries, extracted from the single crystal into the lamellae, were selectively examined in the (010) cross-section by TEM. Figure 2 shows bright-field TEM micrographs and corresponding SAED patterns for Type I (Fig. 2a) and Type II (Fig. 2b) twin boundaries. The two martensite variants are clearly distinguished by contrast, separated by the boundary at approximately 45° to the lamella edges. SAED patterns (zone axis [010]) conform to the expected *a/c*-twinning relationships. The modulation-related satellite reflections occurring within the (hk0) reciprocal plane are not present in the patterns due to the chosen perpendicular ([010] zone axis) orientation.

High-resolution TEM observations are shown in Figure 3, which presents images with increasing detail from left to right. The overall view for Type I twin boundary, Fig. 3a, oriented edge-on, shows no specific features. Higher magnification, Fig. 3b, confirms the atomically sharp boundary following exactly the expected crystallographic orientation (101). The drawn line demonstrates the angle between unit cell diagonals of the two twinned variants and, consequently, also the atomic sharpness of the twin boundary. This angle, in a first approximation disregarding the modulation and monoclinic distortion, is given by tetragonal distortion $c/a$ of the lattice, $\delta = 90 - 2\arctan(c/a) + (\arctan(a/c) - \arctan(c/a)) = 7.53°$. Note that this angle differs from the kink angle on the sample surface, $\alpha_{KINK} = 90 - 2\arctan(c/a) = 3.77°$, see in Fig. 2b. The most atomically detailed analysis, with the position of all atoms assigned to the lattices of the respective twin variants, confirms that the boundary is perfectly atomically sharp. Atomically sharp twin boundaries of compound type were also reported in non-modulated (NM) martensite of $Ni_{53.1}Mn_{25.8}Ga_{21.1}$ [21].

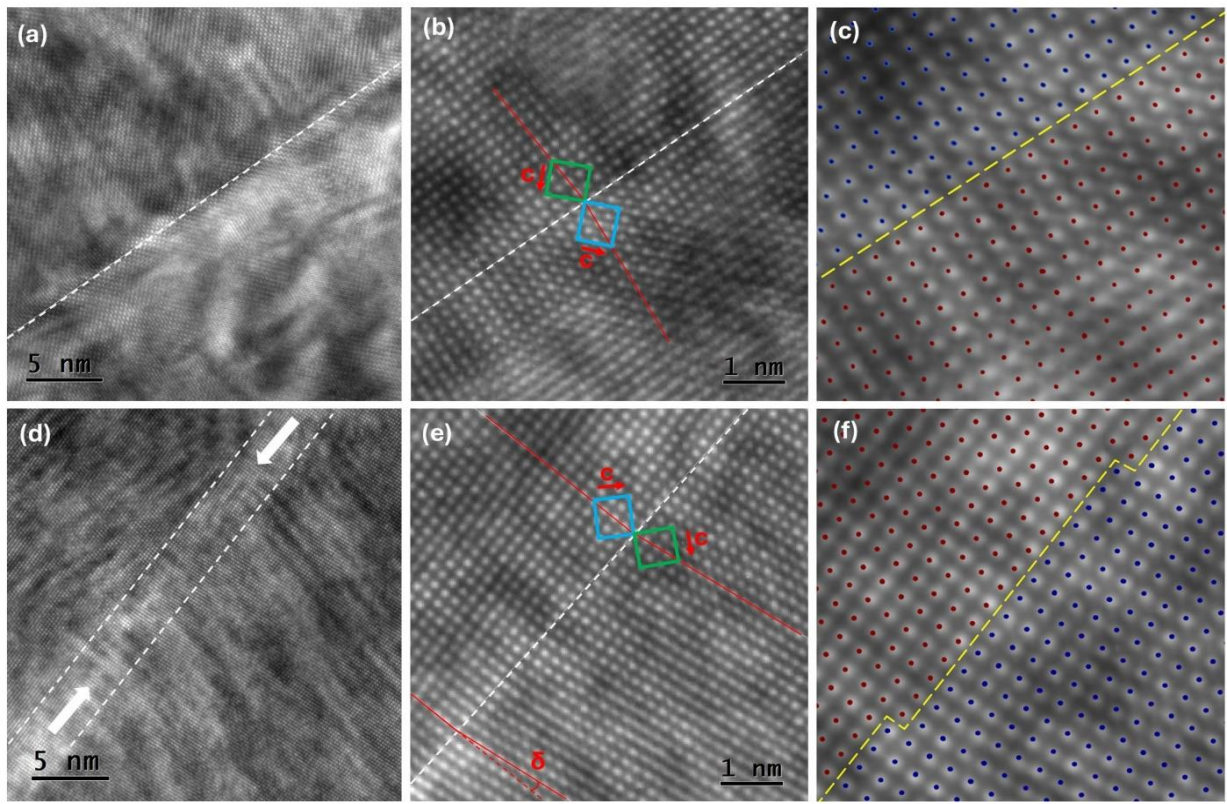

**Figure 3.** HRTEM images illustrating the atomic-scale structure of **(a-c)** Type I and **(d-f)** Type II twin boundaries in 10M Ni–Mn–Ga. Red lines in (b) and (e) are the diagonals of the unit cell forming an angle $\delta = 7.53°$ at the twin boundary.

The corresponding observations for the Type II twin boundary are shown and analyzed in Fig. 3d-e. At low magnification, Fig. 3d, the boundary trace appears to deviate from [101]. A closer look, presented in Fig. 3e, confirms this suggestion. Again, we can draw lines (lattice unit diagonals) through the atomic positions in both variants, but the crossing points of these lines are not on the same line of atoms. Consequently, although the twin boundary locally appears atomically sharp, its trace does not run precisely along the [101] in the studied (010) cut, although the crystallographic theory predicts this.

The boundary appears atomically sharp even at the highest magnification, Fig. 3f. Considering the finite thickness of the foil, about 30 nm, and Type II twin plane deviation from the (101) plane, the atomic sharpness of the boundary comes as a surprise. One can rationalize the observation by the set of regular facets that align all atoms in columns along the electron beam, resulting in a sharp HRTEM image. The existence of an irrational plane demands either bending or facets. The observed sharpness excludes continuous bending, as this would smear the atom traces in the boundary vicinity.

Moreover, assigning all individual atom positions to the lattices of the respective variants reveals steps on the interface (Fig. 3f). These steps cannot be rationalized by the deviation of the Type II twinning plane from the crystallographic (101) plane, because they lie in a perpendicular direction to the tilt axis. One can attribute the observed steps to the twinning dislocations, which can be easily formed in the Type II boundary. These unexpected steps can appear on the twin boundary because of stresses inevitably present in any thin lamellae. Although the stresses are presumed to be similar in both studied lamellae, the steps form only on the Type II boundary, and can be discussed as a possible source of supermobility connected with *a/b*-twin branching [14,22].

Based on our observation and following model suggestions [16,23], we can explain the different mobility of the boundaries. The Type I twin boundary (Fig. 3a-c) aligns with a low-index (101) crystallographic plane (rational K1), consistent with classical twinning theory and topological descriptions involving well-defined disconnections. The existing model for Type I twin boundary mobility suggests that the nucleation of twinning disclinations is difficult, and these may not readily appear on the interface. Hence, the twin boundary of type I appears atomically sharp and straight, with no facets or steps observed.

In contrast, the Type II twin boundary (Fig. 3d-f) nominally lies in an irrational index plane, and our observation suggests the formation of coherent faceting separated by discrete steps. These steps can be associated with misfit disconnections. This faceted structure can be set into motion under low stress, consistent with the high mobility and weak temperature dependence observed in Type II TBs. The supermobility is thus rationalized by the relative ease of step nucleation and motion, coupled with lower energy barriers for twinning due to these coherently faceted local segments.

In any case, compared to usual shape memory materials/martensites, both twin boundaries exhibit extremely low twinning stress and hence high mobility, which may be ascribed to one very particular feature of Ni–Mn–Ga 10M martensite, namely a very low shear elastic constant and extreme elastic anisotropy [12].

A combination of extraordinarily high twin boundary mobility and ferromagnetic ordering is crucial for the magnetoelastic functionality unique for Ni-Mn-Ga. We studied magnetic domain structure across twin boundaries using Lorentz TEM [24]. Figure 4 summarizes the main features observed on both types of boundaries – it presents Fresnel underfocused images and induction colormaps derived from the Transport-of-Intensity Equation (TIE). The micrographs contain typical diffraction artifacts caused by bending contours, which are not discussed further. The "unipolar" line contrast in Fresnel images, which reverses with overfocus and disappears at exact focus, is characteristic of magnetic domain walls (DWs).

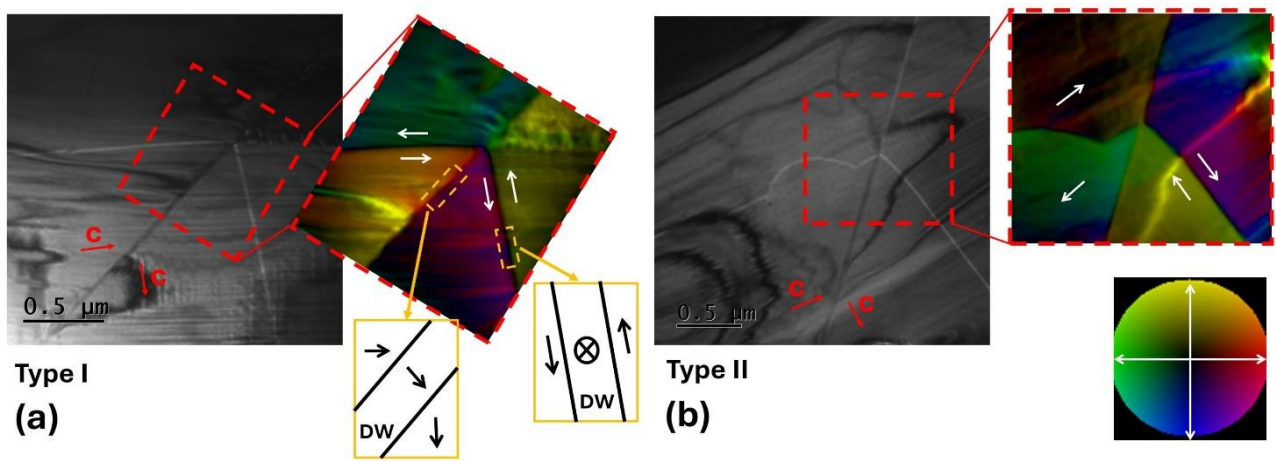

**Figure 4.** Lorentz TEM observation. Fresnel under-focus images with corresponding magnetic induction color maps for **(a)** Type I and **(b)** Type II twin boundary. The magnitude and directions of the local magnetic moments are indicated by the brightness and color phase, referring to the color wheel. The curved 180° domain wall in (b) is attributed to pinning on an antiphase boundary. Overall induction direction is illustrated by white arrows within the maps and small sketches in (a). DW denotes domain walls; the red arrow indicates the orientation of the short *c*-axis, which is an easy magnetization axis.

The twin boundaries are apparent as nearly straight, dark lines running diagonally, Fig. 4a, b. They reverse their contrast upon over-focus/under-focus switching, i.e., they also form a magnetic domain wall. Concerning the domain structure within the twin variants, a roughly horizontal magnetic domain wall changes its orientation to nearly vertical upon crossing the twin boundary from left to right (Fig. 4a). The corresponding DW in the left portion of Fig. 4b bends sharply. This indicates a local pinning on an antiphase boundary [25].

The magnetization distribution in the domains is revealed in detail by the TIE-derived induction maps, which describe the orientation of the magnetic induction in the foil. For both types of twins, the magnetic induction follows broadly the easy magnetization direction. This results in two types of DWs: (i) 90°-like DWs formed within the twin boundary, with in-plane rotations visible as colored DW core, and (ii) straight 180° DWs within the twin variants, recognized by an uncolored (black) core that indicates out-of-plane magnetization reversal. The two types of domain walls are illustrated in the bottom-right in Fig.4a.

Additionally, a segment of the DW within the Type II TB shows contrast reversal in Fresnel underfocus micrograph (Fig. 4b)—from dark at the bottom to bright at the top in Fresnel imaging upon 180° domain crossing. This implies that the magnetization rotates in the opposite direction in the upper part of the lamella compared to the lower part for this segment, i.e., the chirality changes. Such variations in 90° domain switching may stem from the different magnetic properties of the Type II twin boundary itself, or from variations in lamella thickness, which can alter local shape demagnetization effects and consequently the local magnetic energy.

In conclusion, we have characterized the atomic-scale topology of the highly mobile Type I and Type II twin boundaries in 10M Ni-Mn-Ga and observed a connected magnetic domain structure. Our observations provide the first atomically resolved comparison of the topology of both twin boundaries and give insight into the true nature of the highly mobile Type II twin boundary.

Our findings can be summarized as:

i) Type I boundaries are atomically sharp interfaces on a rational (101) plane.

ii) Type II boundaries are intrinsically faceted while also atomically sharp. Our observation, therefore, challenges and effectively rules out the diffuse-interface pictured previously for Type II boundaries.

iii) The magnetic domain structure closely follows the magnetocrystalline anisotropy, with the easy magnetization axis along the *c*-axis. Twin boundaries effectively act as 90°-like magnetic domain walls. In contrast, straight 180° domain walls occur in the individual twin variants and change sharply across both types of twin boundaries, following the change of the easy axis due to twinning.

To further deepen the understanding of the nature of twin boundaries, additional TEM studies with different crystal orientations are needed. This presents a significant challenge; the strong magnetic field in the TEM can easily move the twin boundaries due to the difference in magnetic energy between the twinned variants. Although such work was previously considered nearly impossible, efforts are underway to prepare TEM foils in different orientations to obtain a complete 3D picture of these mobile twin boundaries and its interpretation.

**Acknowledgements**

The authors acknowledge the funding support from the Czech Science Foundation [grant number 24-10334], and the assistance provided by the Ferroic Multifunctionalities project (FerrMion), supported by the Ministry of Education, Youth, and Sports of the Czech Republic [Project No. CZ.02.01.01/00/22_008/0004591], co-funded by the European Union. Moreover, CzechNanoLab project LM2023051 funded by MEYS CR is gratefully acknowledged for the financial support of the measurements/sample fabrication at LNSM Research Infrastructure.

**References**

[1] M. Acet, Ll. Mañosa, A. Planes, Chapter Four - Magnetic-Field-Induced Effects in Martensitic Heusler-Based Magnetic Shape Memory Alloys, in: K.H.J. Buschow (Ed.), Handbook of Magnetic Materials, Elsevier, 2011: pp. 231–289.

[2] I. Aaltio, A. Sozinov, Y. Ge, K. Ullakko, V.K. Lindroos, S.-P. Hannula, Giant Magnetostrictive Materials, Elsevier BV, 2016.

[3] A. Milleret, 4D printing of Ni–Mn–Ga magnetic shape memory alloys: a review, Materials Science and Technology 38 (2022) 593–606.

[4] A. Saren, V. Laitinen, M. Vinogradova, K. Ullakko, Twin boundary mobility in additive manufactured magnetic shape memory alloy 10M Ni-Mn-Ga, Acta Materialia 246 (2023) 118666.

[5] L. Straka, O. Heczko, H. Seiner, N. Lanska, J. Drahokoupil, A. Soroka, S. Fähler, H. Hänninen, A. Sozinov, Highly mobile twinned interface in 10 M modulated Ni–Mn–Ga martensite: Analysis beyond the tetragonal approximation of lattice, Acta Materialia 59 (2011) 7450–7463.

[6] A. Sozinov, N. Lanska, A. Soroka, L. Straka, Highly mobile type II twin boundary in Ni-Mn-Ga five-layered martensite, Applied Physics Letters 99 (2011) 124103.

[7] D. Kellis, A. Smith, K. Ullakko, P. Müllner, Oriented single crystals of Ni–Mn–Ga with very low switching field, Journal of Crystal Growth 359 (2012) 64–68.

[8] L. Straka, A. Soroka, H. Seiner, H. Hänninen, A. Sozinov, Temperature dependence of twinning stress of Type I and Type II twins in 10M modulated Ni–Mn–Ga martensite, Scripta Materialia 67 (2012) 25–28.

[9] O. Heczko, V. Kopecký, A. Sozinov, L. Straka, Magnetic shape memory effect at 1.7 K, Appl. Phys. Lett. 103 (2013) 072405.

[10] D. Musiienko, F. Nilsén, A. Armstrong, M. Rameš, P. Veřtát, R.H. Colman, J. Čapek, P. Müllner, O. Heczko, L. Straka, Effect of crystal quality on twinning stress in Ni–Mn–Ga magnetic shape memory alloys, Journal of Materials Research and Technology 14 (2021) 1934–1944.

[11] H. Seiner, M. Zelený, P. Sedlák, L. Straka, O. Heczko, Experimental Observations versus First-Principles Calculations for Ni–Mn–Ga Ferromagnetic Shape Memory Alloys: A Review, Physica Status Solidi (RRL) – Rapid Research Letters 16 (2022) 2100632.

[12] K. Repček, P. Stoklasová, T. Grabec, P. Sedlák, J. Olejňák, M. Vinogradova, A. Sozinov, P. Veřtát, L. Straka, O. Heczko, H. Seiner, Compliant Lattice Modulations Enable Anomalous Elasticity in Ni–Mn–Ga Martensite, Advanced Materials 36 (2024) 2406672.

[13] O. Heczko, L. Straka, H. Seiner, Different microstructures of mobile twin boundaries in 10M modulated Ni–Mn–Ga martensite, Acta Materialia 61 (2013) 622–631.

[14] H. Seiner, L. Straka, O. Heczko, A microstructural model of motion of macro-twin interfaces in Ni–Mn–Ga 10M martensite, Journal of the Mechanics and Physics of Solids 64 (2014) 198–211.

[15] A.S.K. Mohammed, H. Sehitoglu, Modeling the interface structure of type II twin boundary in B19′ NiTi from an atomistic and topological standpoint, Acta Materialia 183 (2020) 93–109.

[16] D. Shilo, E. Faran, B. Karki, P. Müllner, Twin boundary structure and mobility, Acta Materialia 220 (2021) 117316.

[17] S. Kaufmann, R. Niemann, T. Thersleff, U.K. Rößler, O. Heczko, J. Buschbeck, B. Holzapfel, L. Schultz, S. Fähler, Modulated martensite: why it forms and why it deforms easily, New Journal of Physics 13 (2011) 053029.

[18] M.E. Gruner, S. Fähler, P. Entel, Magnetoelastic coupling and the formation of adaptive martensite in magnetic shape memory alloys, Physica Status Solidi (b) 251 (2014) 2067–2079.

[19]   M. Matsuda, Y. Yasumoto, K. Hashimoto, T. Hara, M. Nishida, Transmission Electron Microscopy of Twins in 10M Martensite in Ni–Mn–Ga Ferromagnetic Shape Memory Alloy, Materials Transactions 53 (2012) 902–906.

[20]   P. Veřtát, H. Seiner, L. Straka, M. Klicpera, A. Sozinov, O. Fabelo, O. Heczko, Hysteretic structural changes within five-layered modulated 10M martensite of Ni–Mn–Ga(–Fe), J. Phys.: Condens. Matter 33 (2021) 265404.

[21]   B. Chen, C. Guan, Y. Li, C. Yang, J. Zhang, G. Liu, L. Li, Y. Peng, Quantitative reorientation behaviors of macro-twin interfaces in shape-memory alloy under compression stimulus *in situ* TEM, Journal of Materials Science & Technology 107 (2022) 243–251.

[22]   O. Heczko, L. Klimša, J. Kopeček, Direct observation of a-b twin laminate in monoclinic five-layered martensite of Ni-Mn-Ga magnetic shape memory single crystal, Scripta Materialia 131 (2017) 76–79.

[23]   B. Karki, P. Müllner, R. Pond, Topological model of *type II* deformation twinning in 10M Ni-Mn-Ga, Acta Materialia 201 (2020) 604–616.

[24]   C. Phatak, A.K. Petford-Long, M. De Graef, Recent advances in Lorentz microscopy, Current Opinion in Solid State and Materials Science 20 (2016) 107–114.

[25]   M. Vronka, L. Straka, M. De Graef, O. Heczko, Antiphase boundaries, magnetic domains, and magnetic vortices in Ni–Mn–Ga single crystals, Acta Materialia 184 (2020) 179–186.